# Searching for an Anomalous $\bar{t}q\gamma$ Coupling Via Single Top Quark Production at a $\gamma\gamma$ Collider

K.J. Abraham[a], K. Whisnant[b], and B.-L. Young[b]

[a] Department of Physics, University of Natal, Pietermaritzburg, SOUTH AFRICA

[b] Department of Physics and Astronomy, Iowa State University, Ames, IA 50011, USA

## Abstract

We investigate the potential of a high-energy $\gamma\gamma$ collider to detect an anomalous $\bar{t}q\gamma$ coupling from observation of the reaction $\gamma\gamma \to t\bar{q}, \bar{t}q$, where $q = c$ or $u$. We find that with $b$-tagging and suitable kinematic cuts this process should be observable if the anomalous coupling $\kappa/\Lambda$ is no less than about $0.05/\text{TeV}$, where $\Lambda$ is the scale of new physics associated with the anomalous interaction. This improves upon the bound possible from observation of top decays at the Tevatron.

## I. INTRODUCTION

Since the discovery of the top quark at the Fermilab Tevatron by the CDF and D0 collaborations [1] there has been much speculation as to whether or not its interactions are in accordance with Standard Model (SM) predictions. Because its mass, around 175 GeV, is of the order of the Fermi scale, the top quark couples quite strongly to the electroweak symmetry-breaking sector. In the minimal SM the electroweak symmetry-breaking sector consists of a single complex fundamental Higgs scalar, but "triviality" [2] and "naturalness" [3] of the scalar sector suggest that in fact the Higgs sector, and therefore the top quark mass generation mechanism, may be more complicated. It is therefore plausible to assume that the Higgs sector of the SM is an effective theory, and that new physics phenomena may be manifested through effective interactions of the top quark [4].

One interesting sub-set of effective top quark interactions mediates flavour changing neutral $t$ decays, i.e., $t \to cZ$, $cg$, and $c\gamma$. The SM predictions for the corresponding branching fractions are unobservably small [5]; thus any experimental evidence for such decays will be an unambiguous signal for physics beyond the Standard Model. Furthermore, it has been argued that these decay rates may be enhanced significantly in many extensions of the standard model, such as SUSY or other models with multiple Higgs doublets [5,6], models with new dynamical interactions of the top quark [7], and models where the top quark has a composite [8] or soliton structure [9].

Many aspects of such anomalous top-quark couplings have already been investigated in hadron and lepton colliders. These couplings give contributions to low-energy observables such as the partial width ratio $R_b = \Gamma(Z \to b\bar{b})/\Gamma(Z \to \text{hadrons})$ measured at LEP-I [10], or the branching fraction for $b \to s\gamma$ [11]. Other constraints from low-energy processes on anomalous couplings of top quark have also been considered in the literature [12]. Furthermore, such couplings would also affect top-quark production and decay processes at hadron and $e^+e^-$ colliders [13,14]. In Ref. [15], the experimental constraints on an anomalous top-quark coupling $\bar{t}cZ$ and the experimental observability of the induced rare decay mode $t \to Zc$, at the Fermilab Tevatron and the CERN LHC, have been investigated in detail. The observability of an anomalous coupling $\bar{t}cg$ has been studied for $t \to cg$ decays at the Tevatron [16], for single top production in association with a charm quark [17,18], and for direct top production, $gq \to t$ ($q = c$ or $u$) at the Tevatron and the LHC [19].

In this paper, we examine the possibility of searching for the anomalous top-quark couplings $\bar{t}c\gamma$ and $\bar{t}u\gamma$ at a high-energy $\gamma\gamma$ collider. Such a collider may be constructed by the compton scattering of laser light off the $e^+$ and $e^-$ beams in an $e^+e^-$ linear collider [20]. We will consider the anomalous effective Lagrangian which includes only the lowest dimension,

$CP$-conserving operators which give rise to anomalous $\bar{t}q\gamma$ vertices, namely

$$\Delta \mathcal{L}^{eff} = \frac{e}{\Lambda}[\kappa_c \bar{t}\sigma_{\mu\nu}c + \kappa_u \bar{t}\sigma_{\mu\nu}u]F^{\mu\nu} + h.c., \quad (1)$$

where $F^{\mu\nu}$ is the electromagnetic field strength tensor, $e$ is the electromagnetic coupling constant, $\Lambda$ is the cutoff of the effective theory, which is generally taken to be the order of 1 TeV, and the parameters $\kappa_c$ and $\kappa_u$ can be interpreted as the strengths of the anomalous interactions. Here we do not choose a particular scale, and consider only the ratio $\kappa/\Lambda$. In principle, one could consider a more complex form factor with the tensor structure $\sigma_{\mu\nu}(A + B\gamma^5)$, however, if we consider only polarization averaged cross-sections, we may set $B = 0$ without any loss of generality in probing the strength of such couplings.

The constraint from the inclusive branching ratio for the process $b \to s\gamma$ [21] on the anomalous top-quark coupling $\bar{t}c\gamma$ gives $\kappa_c/\Lambda < 0.16$/TeV in the absence of an anomalous $\bar{t}cg$ coupling [23] (the limit with a non-zero $\bar{t}cg$ coupling is 0.28/TeV). The analysis can be extended to the $\bar{t}u\gamma$ coupling by realizing that the anomalous contribution to $b \to s\gamma$ will be suppressed by the CKM factor $V_{us} \approx 0.22$ compared to the $\bar{t}c\gamma$ case, which translates into the limit $\kappa_u/\Lambda < 0.72$/TeV. The current direct experimental bound comes from CDF data on the top branching fractions to a jet plus photon [22]

$$BF(t \to c\gamma) + BF(t \to u\gamma) < 2.9\%, \quad (2)$$

at 95% Confidence Level (CL), which translates into the limit

$$\kappa/\Lambda \equiv \sqrt{\kappa_c^2 + \kappa_u^2}/\Lambda < (0.73/\text{TeV})/\sqrt{BF(t \to bW)}. \quad (3)$$

This experimental limit therefore does not improve on the $b \to s\gamma$ constraint in either case, although the bounds on $\kappa_u$ are numerically about the same.

By searching for the decay $t \to q\gamma$ in $t\bar{t}$ production [23] one can potentially observe the anomalous couplings down to $\kappa/\Lambda = 0.12$/TeV at the Tevatron Upgrade with 10 fb$^{-1}$ of integrated luminosity and to $\kappa/\Lambda = 0.01$/TeV at the LHC with 100 fb$^{-1}$ of integrated luminosity, assuming that the light quark jet can not be identified. We will find that the $\gamma\gamma$ collider may be able to improve upon the potential Tevatron limit on these anomalous couplings by roughly a factor of 2.5. The relatively clean environment of a $\gamma\gamma$ collider presents an opportunity for differentiating between the $\bar{t}c\gamma$ and $\bar{t}u\gamma$ couplings, assuming $c$-tagging by the identification of D mesons produced by hadronization, such as is done in various $e^+e^-$ experiments [24], is possible. This particular option does not seem to be feasible at hadron colliders.

## II. ANOMALOUS SINGLE TOP PRODUCTION AT A $\gamma\gamma$ COLLIDER

The anomalous vertices under consideration here lead to the interaction $\gamma\gamma \to t\bar{q}$, where $q = c$ or $u$. There are four diagrams which contribute, each with one anomalous vertex and one SM vertex: either the top quark or light quark can be exchanged in the $t$ or $u$ channel. For very high energies, in the limit that both fermion masses can be ignored, the total cross section approaches $\sigma(\gamma\gamma \to t\bar{q}) = 64\pi\alpha^2(\kappa/\Lambda)^2$. This result is finite despite the $t$ and $u$-channel poles because the momentum coupling in Eq. 1 regulates the divergence when the fermion masses vanish. For $\kappa/\Lambda = 0.16/\text{TeV}$, the maximal value allowed for the $\bar{t}c\gamma$ coupling in the absence of a $\bar{t}qg$ coupling, this gives about 120 fb (for the corresponding $\bar{t}u\gamma$ coupling this is increased by roughly a factor of 20). In this paper we will use $\kappa/\Lambda = 0.16/\text{TeV}$ in our calculations of the signal unless noted otherwise. The cross section for single antitop production $\gamma\gamma \to \bar{t}q$ is the same; henceforth in this paper all signal rates and discussions will include the sum of the single top and antitop signals. The cleanest signal occurs when the top quark decays semileptonically, which gives the signature $bj\ell\displaystyle{\not}p_T$, where $j$ is a light quark jet and $\ell = e$ or $\mu$.

Since a likely $\gamma\gamma$ collider will have a maximum CM energy of 500 GeV to 1 TeV [25], the top mass cannot be completely ignored, and we have calculated the full matrix element for the $2 \to 2$ process, assuming the top quark is on shell. Because various cuts must be introduced to simulate a detector and to eliminate backgrounds, we have then calculated the top quark decay, using the exact matrix element for the top 3-body semileptonic decay, assuming an on-shell $W$. Because these couplings do not favor one helicity, the top quark is produced unpolarized and we therefore do not have to worry about spin correlations between the top production and decay. The cross sections were calculated via a Monte Carlo program, using both helicity amplitudes and Dirac matrices, and agreement was found to within 1% between the two methods. We have also assumed that $BF(t \to bW) \approx 1$, which would be the case if there were no non-standard top decays other than $t \to q\gamma$. If there are non-standard top decays with a significant branching fraction, the signal results quoted in this paper would be reduced at most by a factor of two [23]. For $\sqrt{s} = 500$ GeV and $m_t = 175$ GeV we find a total signal cross section of

$$\sigma(\gamma\gamma \to t\bar{q} + \bar{t}q \to bc\ell\nu) = 76.4 \text{ fb} \left(\frac{\kappa/\Lambda}{0.16/\text{TeV}}\right)^2, \tag{4}$$

which for a $\bar{t}c\gamma$ ($\bar{t}u\gamma$) coupling of $\kappa/\Lambda = 0.16/\text{TeV}$ ($0.72/\text{TeV}$) translates into 764 (15280) events with an integrated luminosity of 10 fb$^{-1}$. Therefore we see that this is a viable signal.

## III. ACCEPTANCE CUTS AND BACKGROUNDS

There are several potentially severe SM backgrounds to the signal. The largest is $\gamma\gamma \to W^+W^-$, which has a cross section of about 88 pb at $\sqrt{s} = 500$ GeV [26]. This will mimic the signal when one $W$ decays leptonically and the other decays into two light quark jets, giving a cross section of about 26 pb. To make a quantitative analysis of the experimental sensitivity to the anomalous couplings, we have done a monte carlo study of $\gamma\gamma \to W^+W^- \to \ell\nu q'\bar{q}''$ using the full helicity amplitudes [30], imposing the following basic acceptance cuts

$$p_T^\ell > 15 \text{ GeV}, \qquad p_T^j > 15 \text{ GeV}, \qquad E_T^{miss} > 15 \text{ GeV},$$
$$|\eta^\ell|, |\eta^j| < 1.5, \qquad \Delta R_{\ell j}, \Delta R_{jj} > 0.4, \tag{5}$$

where $p_T$ denotes transverse momentum, $\eta$ denotes pseudo-rapidity, and $\Delta R$ denotes the separation in the azimuthal angle-pseudo rapidity plane. With these basic cuts, the signal rate is reduced to about 33 fb, while the $W^+W^-$ background is now about 4.8 pb.

A significant reduction in the background is possible if $b$-tagging is employed. We assume a 50% efficiency for detecting a $b$-quark jet, and a 0.4% mis-tagging rate where a light quark jet is misidentified as a $b$-quark [28,29]. The signal is then halved, while the $W^+W^-$ background is reduced a level comparable with the signal.

Next we can then use the fact that for this background the two jets should reconstruct to the $W$ mass, while for the signal the $m(b\bar{c})$ invariant mass distribution is peaked towards its maximal value of $\sqrt{s} - m_W$=420 GeV, which can be understood by realizing that the charm jet and the $b$ jet from top decay tend to be back-to-back. If we exclude values of $m(jj)$ which include the $W$ resonance, we can substantially reduce this background. To quantify this in our calculation, we assume a Gaussian energy smearing for the electromagnetic and hadronic calorimetry as follows

$$\Delta E/E = 30\%/\sqrt{E} \oplus 1\%, \quad \text{for lepton and photon}$$
$$= 80\%/\sqrt{E} \oplus 5\%, \quad \text{for jets,} \tag{6}$$

where the $\oplus$ indicates that the $E$-dependent and $E$-independent errors are to be added in quadrature, and $E$ is measured in GeV. If we impose the cut

$$m(jj) > 106 \text{ GeV}, \tag{7}$$

then we find that the resonant contributions are reduced to $\mathcal{O}(10^{-2})$ fb, effectively eliminating this background. The signal is affected very little by the $m(jj)$ cut, as can be seen from Fig. 1.

In addition to resonant $W^+W^-$ production, there is a set of Feynman graphs contributing to nonresonant $\gamma\gamma \to \ell\nu q'\bar{q}$ production, where the invariant mass of the lepton and quark pairs are not both at the $W$ mass peak. Due to the large number of different graphs involved, a precise analytical treatment is beyond the scope of this letter, and indeed has not yet been carried out. However, we will make use of the results for resonant and nonresonant contributions analyzed numerically in Ref. [27] to show how this background can be suppressed. Most of the nonresonant cross section occurs when one pair of fermions is on a $W$ peak and one of the two other fermions is roughly collinear with an incoming photon. The pseudo-rapidity cut in Eq. 6, larger than likely required from detector geometry, has been chosen to reduce these forward contributions. From Ref. [27] the cross section for $\gamma\gamma \to \ell\nu q'\bar{q}''$ at $\sqrt{s} = 500$ GeV, excluding the $W^+W^-$ resonant contributions, is about 67 fb after the $\eta$ cuts (which corresponds to cuts on the charged particles of about $|\cos\theta| < 0.92$ in the COM frame of the incoming photons). We then multiplied this result by 8 to account for two generations each of leptons and quarks and for positive and negative charged leptons in the final state, and divided by 2 since we expect that in roughly half of these events the $W$ has on-shell decays into quarks, which will be eliminated by the $m(jj)$ cut. The net result is a $Wjj \to \ell\nu jj$ background of about 270 fb, which will be reduced by more than two orders of magnitude by $b$-tagging alone.

A further reduction of the nonresonant background can be made by determining the invariant mass $m(bW)$, which for the signal should be strongly peaked near $m_t$ and for the background should be much flatter. In order to do this reconstruction one must determine the $W$ momentum. The neutrino transverse momentum can be inferred from the missing $p_T$, but its longitudinal momentum is undetermined since in general the exact initial photon energies may not be known. However, if we assume that the $W$ which decays leptonically is on mass shell, one can determine the neutrino longitudinal momentum up to a two-fold ambiguity [31]. We can then take the solution which gives $m(bW)$ closest to $m_t$, where the $b$ quark is identified via a $b$-tag. Applying the cut

$$|m(bW) - m_t| < 30 \text{ GeV}, \qquad (8)$$

will then provide a strong constraint on the background, while the signal is generally reduced by only about 10-20% [16,23].

To simulate the effect of the remaining cuts (in $p_T$, $\Delta R$, $m(jj)$, and $m(bW)$) on the nonresonant background, we have assumed that the distributions that survive the $\eta$ cuts are relatively flat. This assumption is reasonable as the region close to the beam pipe, where the bulk of the non-resonant cross-section arises [27], has been excluded. The fraction of the non-resonant cross-section which survives is then proportional to the volume of phase space allowed after our additional cuts, which is reduced by a factor of 7. If we assume that the cross section is correspondingly reduced, we get a nonresonant background of about .16 fb. The successive effect of all the cuts on the signal and primary backgrounds are summarized in Table 1.

There are also SM final states $bc\ell^{\pm}\nu$ from both resonant and nonresonant backgrounds. However, these processes are suppressed by a factor $|V_{bc}|^2 \approx 1/400$ compared to the final states $ud\ell^{\pm}\nu$ and $cs\ell^{\pm}\nu$. Although they are reduced only by a factor of 2 when $b$-tagging is employed, they are still more than a factor of 3 smaller than the corresponding processes with light quarks after $b$-tagging. We can account for these backgrounds by multiplying the light quark background cross section by 21/16, which gives an overall estimated background at the 0.2 pb level, which corresponds to 2 events for 10 fb$^{-1}$ integrated luminosity.

The signal cross section after all cuts is about 23.4 fb for $\kappa/\Lambda = 0.16$/TeV. However, about half the signal is lost due to the requirement of $b$-tagging, still leaving a cross-section large enough to be phenomenologically viable, assuming $\mathcal{L} = 10^{-1}$ fb. At realistic $\gamma\gamma$ colliders, of course, $\sqrt{s}$ is not 500 GeV as we assumed above, but lower; taking $\sqrt{s} = 400$ GeV, as for example in [26], leads to a reduction of $\sim 10\%$ in the signal cross-section and should have no significant influence on possible discovery bounds.

## IV. ANOMALOUS COUPLING LIMITS AND DISCUSSION

To estimate the sensitivity to the anomalous couplings for a given integrated luminosity, we require that the signal be observed at the 3-$\sigma$ level,

$$S \geq 3\sqrt{S+B}, \tag{9}$$

where $S$ and $B$ are the number of signal and background events, respectively, after all cuts are made. In our case ($B \approx 2$), this corresponds to at least 11 signal events. For $\sqrt{s} = 500$ GeV and an integrated luminosity of 10 fb$^{-1}$, the discovery limit for $\kappa/\Lambda$ is then about 0.048/TeV. For a more realistic CM energy of $\sqrt{s} = 400$ GeV, the discovery limit can be written approximately as

$$\kappa/\Lambda \leq \frac{0.051/\text{TeV}}{\sqrt{\mathcal{L}/10 \text{ fb}^{-1}}}, \qquad (10)$$

when the integrated luminosity $\mathcal{L}$ lies in the range 5-20 fb$^{-1}$. A $\gamma\gamma$ collider with maximum CM energy of 1 TeV and an effective $\sqrt{s} = 800$ GeV may be able to reduce this by about 10%. We note that if the background has been underestimated by a factor of 3, these discovery limits are raised by only about 10%.

We have also examined the possibility of looking for $e^+e^- \to t\bar{q}+\bar{t}q$ in an electron-positron collider. However, the production cross section is much lower than for a $\gamma\gamma$ collider. We find that for $\kappa/\Lambda = 0.16/\text{TeV}$ and $m_t = 175$ GeV the total cross section for $e^+e^- \to t\bar{q} + \bar{t}q$ before cuts is about 1.8 fb at $\sqrt{s} = 200$ GeV and 4.4 fb at $\sqrt{s} = 500$ GeV. These are more than an order of magnitude smaller than the corresponding $\gamma\gamma$ cross sections, and means that an $e^+e^-$ collider cannot effectively probe the $\bar{t}q\gamma$ coupling.

In summary, we have shown that a $\gamma\gamma$ collider with $\sqrt{s} = 500$ GeV can probe an anomalous $\bar{t}c\gamma$ or $\bar{t}u\gamma$ coupling down to the level of $\kappa/\Lambda \sim 0.05/\text{TeV}$ for the integrated luminosities expected at such machines. As noted previously, a reduction in $\sqrt{s}$ to 400 GeV does not significantly affect this limit. This will be much more sensitive than looking for $e^+e^- \to t\bar{c}$ at a similar energy, and is a factor of about 2.5 better than the limit expected to be obtained from studying top decays at the upgraded Tevatron, where backgrounds are not so easy to suppress. If charm tagging is available it also offers the possibility of distinguishing between the anomalous $\bar{t}c\gamma$ and $\bar{t}u\gamma$ couplings, which plausibly could be quite different.

## V. ACKNOWLEDGMENTS

This work was supported in part by the U.S. Department of Energy under Contract DE-FG02-94ER40817 (KW and BLY). KJA wishes to acknowledge the generous support of IITAP during the course of this investigation.

TABLES

TABLE I. Cross sections in units of fb for the $t\bar{q}, \bar{t}q \to bq\ell^{\pm}\not{p}_T$ signal with $\kappa/\Lambda = 0.16/TeV$ and the SM backgrounds. The successive effect of the cuts on the signal and $W^+W^-$ background have been explicitly calculated, while the effects on the nonresonant background values have been estimated as discussed in the text. The dashes indicate cross sections too small to be of interest.

| (a). Cuts | signal $t\bar{c}, \bar{t}c \to bc\ell^{\pm}\nu$ | $WW \to jj\ell^{\pm}\nu$ | $Wjj \to jj\ell^{\pm}\nu$ |
|---|---|---|---|
| none | 76.4 | 26000 | 2140 |
| $\eta$ only | 38.2 | 6928 | 270 |
| basic+smear | 33.5 | 4786 | 175 |
| b-tag | 16.75 | 19.0 | 0.70 |
| m(jj) | 15.7 | - | 0.65 |
| m(bW) | 11.7 | - | 0.16 |

FIGURE CAPTIONS

FIG. 1 The $\gamma\gamma \to t\bar{q} + \bar{t}q \to bq\ell^{\pm}\not{p}_T$ cross section versus the jet-jet invariant mass $m(jj)$ at $\sqrt{s} = 500$ GeV after the basic cuts have been implemented. The solid (dotted) histogram corresponds to the result before (after) the effects of detector smearing are included.

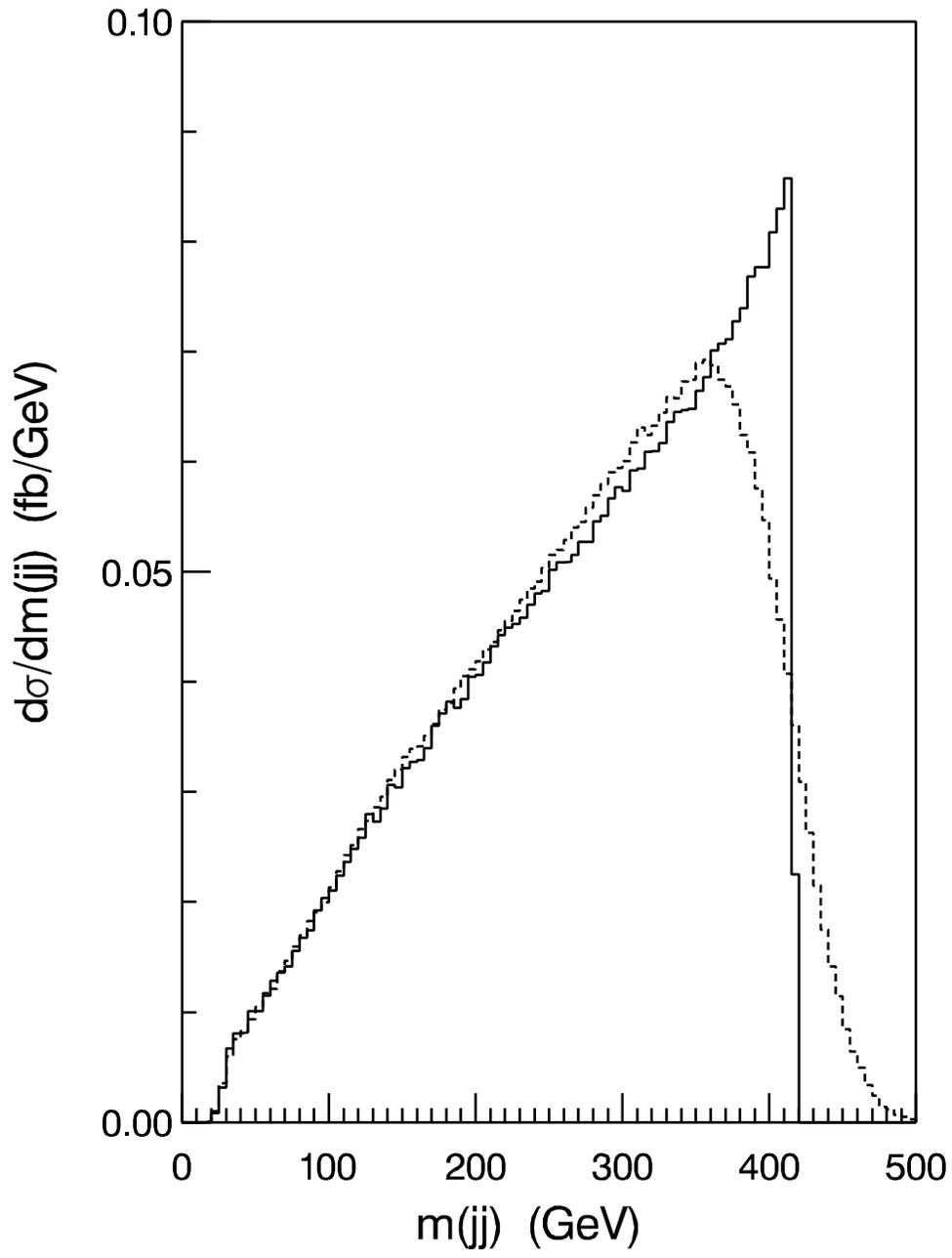

Fig. 1